# FAIRNESS- AND UNCERTAINTY-AWARE DATA GENERATION FOR DATA-DRIVEN DESIGN

**Jiarui Xie**
McGill University
Montreal, QC,
Canada

**Chonghui Zhang**
McGill University
Montreal, QC,
Canada

**Lijun Sun**
McGill University
Montreal, QC,
Canada

**Yaoyao Fiona Zhao***
McGill University
Montreal, QC,
Canada

## ABSTRACT

*The design dataset is the backbone of data-driven design. Ideally, the dataset should be fairly distributed in both shape and property spaces to efficiently explore the underlying relationship. However, the classical experimental design focuses on shape diversity and thus yields biased exploration in the property space. Recently developed methods either conduct subset selection from a large dataset or employ assumptions with severe limitations. In this paper, fairness- and uncertainty-aware data generation (FairGen) is proposed to actively detect and generate missing properties starting from a small dataset. At each iteration, its coverage module computes the data coverage to guide the selection of the target properties. The uncertainty module ensures that the generative model can make certain and thus accurate shape predictions. Integrating the two modules, Bayesian optimization determines the target properties, which are thereafter fed into the generative model to predict the associated shapes. The new designs, whose properties are analyzed by simulation, are added to the design dataset. An S-slot design dataset case study was implemented to demonstrate the efficiency of FairGen in auxetic structural design. Compared with grid and randomized sampling, FairGen increased the coverage score at twice the speed and significantly expanded the sampled region in the property space. As a result, the generative models trained with FairGen-generated datasets showed consistent and significant reductions in mean absolute errors.*

Keywords: machine learning; data-driven design; fairness and diversity; uncertainty; data generation; adaptive sampling.

## 1. INTRODUCTION

Design space exploration (DSE) searches through a wide range of design parameters and configurations for optimal engineering design solutions [1, 2]. With the advent of advanced machine learning (ML) algorithms, data-driven design methods have emerged and allowed rapid, accurate and cost-efficient design generation and DSE [3]. In mechanical design, various data-driven design pipelines and databases have been constructed to aid design tasks such as metamaterial and structural design [4-6].

Conventional data-driven design (Figure 1) starts with the parameterization of target designs, followed by design of experiments (DOE) techniques that sample from the design space such as geometric space [4]. Recently, non-parametric representations such as topology optimization have been implemented in data-driven and generative design [7, 8]. With design representations and experimental plans established, designs can be generated in computer-aided design environments. Thereafter, the mechanical and physical properties of the designs can be analyzed using simulation or real-world experiments. After the data are acquired from the experiments, the relationship between the design space and the property space can be modeled using ML. There are typically two modeling tasks: design performance prediction and generative models. Performance prediction models predict the properties of a design given the design parameters. They are frequently used as surrogate models to replace computationally heavy simulations and speed up design optimization [9]. Generative models, characterizing the inverse relationship, generate designs with respect to specified properties or constraints [8]. It is more difficult to learn such one-to-many relationships that one input could correspond to multiple outputs [10]. Although such workflows have been commonly implemented and have been contributing to various design research discoveries, risks of representation bias stemming from data acquisition might cause fairness issues in the dataset and thus compromise the performance of ML models.

Representation bias describes the phenomenon that some parts of the target population are underrepresented in the dataset [11]. In design datasets, the most salient representation bias resides in the property space, where samples are passively populated [12]. DOE conducted on the design space ensures the generation of diverse design geometries and configurations. Nonetheless, it results in skewed underlying distribution in the property space due to the nonlinear relationship between design shape and properties. Consequently, design datasets are commonly unbalanced in the property space with intensively searched regions, voids in the sampled regions, and unexplored regions [13]. Representation bias in the dataset will propagate to the ML models and eventually yields unsatisfactory designs. Unexplored regions imply missing knowledge in the dataset and contribute to inaccurate predictions of unexplored properties. Data imbalance may cause the ML model to focus on the intensively sampled property regions, while overlooking the underrepresented properties.

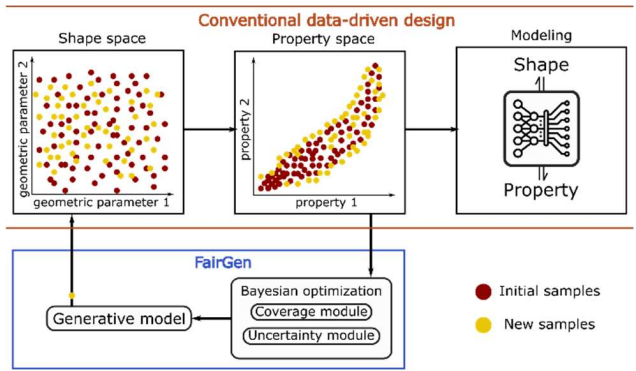

Figure 1: Schematics of the procedures in data-driven design and the role of FairGen.

Current methods to mitigate representation bias in design datasets mainly concentrate on data augmentation such as over-sampling and under-sampling. Over-sampling techniques increase the sample size by partially altering existing samples or generating new synthetic samples [14]. Down-sampling removes similar samples from the overrepresented groups [15]. However, the former might contribute to the overfitting of existing samples and the latter might remove samples with important information [16]. Chan et al. [13] proposed METASET to select an unbiased subset from a large metamaterial shape database. Determinantal point process (DPP) is utilized to model the diversity in both shape and property spaces, which are jointly considered to evaluate the subsets. The selected subset is highly diverse with a small sample size, offering better predictive performance and less training time. Lee et al. [12] proposed t-METASET that iteratively generates diverse unit cell shapes and acquires diverse properties from the existing samples. Its task-aware functionality guides property sampling toward the designated region. However, these methods only implement subset selection in the property space and thus cannot actively expand the sampled region. Wang et al. [6] designed a shape perturbation algorithm that gradually samples new properties toward the unexplored regions in the property space. It builds on the assumption that a small perturbation in the shape of a design will yield a small change in its properties. The rarely explored regions can be populated by slightly altering the shapes of the existing samples nearby. However, the assumption has serious limitations because small perturbations in different shapes can yield considerably different property shifts, which potentially makes the process uncontrollable and inefficient.

To ensure fair and efficient property space exploration, there needs to be a more reliable method that detects the regions where the existing dataset has insufficient coverage and accurately generates designs to increase coverage. Accurate design generation requires this method to model the relationship between shapes and properties instead of relying on assumptions such as small perturbation. Generative models and reinforcement learning (RL) have recently been implemented to generate design geometries that can achieve desirable properties. Chen and Ahmed [3] presented performance augmented diverse generative adversarial network (GAN) that combines GAN loss with performance augmented DPP loss in the training process. Such a GAN model learns to synthesize the training design data while generating diverse shapes with desirable properties. Considering there are usually multiple target properties in design tasks, Chen and Ahmed [17] integrated performance augmented diverse GAN with multi-objective Bayesian optimization (BO). As demonstrated in the case studies, this pipeline can generate diverse shapes and facilitate the exploration of the full Pareto fronts in the property space. Nobari et al. [18] proposed performance conditional diverse GAN to enable the generation of designs with designated properties. Compared with performance augmented diverse GAN, this model is more flexible as the users can appoint desirable properties instead of maximizing or minimizing properties. Instead of GANs that directly generate designs with desirable properties, RL traverses the property space and moves toward the optimal properties iteratively. Jang et al. [4] trained an RL agent that iteratively generates diverse designs by rewarding the diversity of topology. Compared with conventional greedy search, this method can generate 5% more design shapes on average in the tire design case study. Agrawal and McComb [5] trained an RL-based design agent that explores the design space with varying model fidelity. This framework is computationally efficient because of its embedded mechanism to tradeoff between low- and high-fidelity models during DSE.

The common limitation of the above generative models and RL pipelines is that a specific application must be defined before the initiation of DSE. The goals of these methods are to find the optimal or designated properties within the design space. It is straightforward to define the optimality of some properties such as tensile strength, whose optimality means its maximum. For properties such as elastic modulus (EM) and porosity, optimality is dependent on the use case. For instance, soft robotics would favor designs with relatively small EM, while the EM of human bone implants should be close to the EM of original human bones for improved bone integration. To prepare a general-purpose database for various applications, there needs to be a method that fairly explores the property space with no optimal properties specified. This method can explore and exploit the potential of a type of design to facilitate future DSE and design decision-making.

Adaptive sampling is an efficient method to actively generate new data from insufficiently regions [19]. Typical adaptive sampling techniques select new samples according to the predictive performance or uncertainty of ML models. When determining the new samples using predictive performance, the feature space is usually segregated into subspaces. Based on the test set, the subspaces that exhibit the highest predictive error will be the regions of interest (ROI) for adaptive sampling. For instance, Zhang et al. [20] designed an adaptive sampling technique to iteratively improve the performance of surrogate models in design optimization. The test set is divided into subgroups using K-means clustering and KNN. The subgroup that possesses the highest total prediction error is the ROI. Thereafter, maximum curvature is used to select a set of points from the ROI to generate new samples. Adaptive sampling based on predictive performance has also been implemented for structural design [21], configuration design [22], electromagnetic design [23], and protective coating design [24]. Uncertainty metrics such as entropy of prediction probabilities are also widely deployed in adaptive sampling. Gaussian process regression models are trained as surrogate models in various design optimization works and can guide adaptive sampling because of their inherent uncertainty measurement functionality [19]. Xu et al. [25] and Liu et al. [26] implemented Gaussian process adaptive sampling for hall effect sensor design optimization and functionally graded cellular structure design optimization, respectively. Nonetheless, the existing adaptive sampling methods lack the ability to deal with inverse problems and one-to-many relationships in generative design.

In this paper, the authors propose a fairness- and uncertainty-aware data generation (FairGen) pipeline that adaptively samples designs with missing properties (Figure 1). It adds an iterative process to the conventional design pipeline to fairly generate new samples. FairGen does not only exploit the voids within the sampled region, but also gradually expands the sampled region to explore the property space. The key contributions and features of this pipeline include:

- Introducing a fairness metric to design data generation to quantify and visualize data coverage.
- Constructing a novel pipeline and generative models to directly generate missing properties in the dataset.
- Building deep ensemble to model the predictive uncertainties of the generative models and guide the generative process.
- Proposing a pipeline to achieve adaptive sampling for data-driven design problems with inverse modeling and one-to-many relationships.
- FairGen rapidly explores the property space to expand the sampled regions.
- FairGen significantly improves the performance of inverse design models.

The remainder of this paper is organized as follows. Section 2 introduces the methodology of FairGen, including the formulation of the coverage and uncertainty modules. Section 3 presents the setting and procedures of the S-slot auxetic design property space exploration case study. Section 4 discusses the results with respect to the coverage increase rate, property space sampled region expansion, and the impact on generative models. Section 5 highlights the remarks of this research.

## 2. METHODOLOGY

This section illustrates the elements and procedures of FairGen. Section 2.1 demonstrates the FairGen pipeline. Section 2.2 discusses the coverage module with respect to data fairness, data coverage, and Voronoi diagram to construct the coverage map. Section 2.3 discusses the uncertainty module with respect to mixture density network (MDN) and deep ensemble method to capture the predictive uncertainty. Section 2.4 discusses BO integrating the coverage and uncertainty modules to find the target properties.

### 2.1. FairGen pipeline

Figure 2 visualizes the pipeline and modules of FairGen. This pipeline starts with an initial dataset ($D^0$) sampled from the shape space ($R^d$) that contains d geometric parameters. The p types of properties of the n designs from the $D^0$ are analyzed using simulation, then populated in the property space, $R^p$. At each iteration, the mission is to find the empty regions in the property space and generate designs to supplement them. Thus, a data coverage module is built to indicate the uncovered regions. Due to the limitation of the existing knowledge, it is infeasible to accurately generate all missing properties at once. This becomes an optimization problem in which an optimal set of $n_p$ target property samples ($D^P$) is searched. Thus, BO is implemented at every iteration to find a solution of $D^P$ that maximally increases the data coverage in the property space. The coverage module computes the covered area as the coverage score ($S_C$) when $D^P$ is added to the existing dataset. After $D^P$ is solved by BO, the corresponding shape sets ($D^S$) that can provide $D^P$ must be found.

MDN, a generative model, is trained using the existing dataset and predicts the shapes given $D^P$. However, BO purely maximizing the coverage score will yield target properties that maximize the covered area and thus is far away from the existing samples. MDN trained on the existing samples will generate inaccurate shape predictions that do not correspond to the target properties. This raises a conflict between the expansion of coverage and the predictive performance of the generative model. Therefore, an uncertainty module consisting of multiple MDNs is established to compute the predictive uncertainties regarding $D^P$. An uncertainty score $S_U$ characterizing the predictive uncertainties is added to the objective function as a trade-off with the coverage score. This way, BO is encouraged to find a $D^P$ that both efficiently increases the data coverage and ensures accurate shape prediction. The shapes predicted by the MDNs from the uncertainty module are analyzed in simulation to find the actual properties. The new shape-property set is added to the existing dataset, which forms a new dataset $D^i$, where i is the number of iterations. This pipeline can be executed iteratively until the desired $S_C$ is reached or the designated computational resource is exhausted.

## 2.2. Coverage module

The first task of measuring representation bias is to establish a metric. There are two mentalities to model representation bias: fairness and diversity. Fairness describes the lack of bias and diversity describes the richness of variety [27]. Distance-based diversity metrics have been commonly implemented in the research domain of data-driven design [12, 13, 17, 18]. For example, Chan et al. [13] implemented DPP where Euclidean distance and Hausdorff distance were used to construct similarity kernels for 2-dimensional and 3-dimensional shapes, respectively. The authors argued that diversity metrics such as DPP are more flexible and easily implementable to be incorporated into ML pipelines. However, it is hard to use diversity metrics to quantify and visualize data coverage. The quantification and visualization of data coverage at different sample sizes and different $D^P$'s help evaluate and guide the data generation process; thus, a data coverage module must be constructed with a suitable fairness metric.

Asudeh et al. [28] defined the notion of coverage of a point in a continuous-valued feature space. Given a dataset D, a query point q, a distance function $\Delta$, a vicinity value $\rho$, and a threshold value k, the coverage of q by D is defined:

$$Cov_{\rho,k}(q, D) = \begin{cases} true & if\ |\{t \in D | \Delta(t, q) \leq \rho\}| \geq k \\ false & otherwise \end{cases} \quad (1)$$

This definition essentially checks if the query point is at the vicinity defined by $\rho$ and $\Delta$ of at least k data points from the dataset D. With user-defined $\rho$ and k, a region covered by the dataset can be computed by:

$$S_C(D) = \{q | Cov(q, D) = True\} \quad (2)$$

In FairGen, the coverage of the property space is to be improved. The covered area is the coverage score of the coverage module to quantify coverage and evaluate the selection of target properties. BO will utilize the coverage score to find a set of target properties that optimally increases the improvement of data coverage. The definition of data coverage is clear and straightforward to understand and implement. The covered region can also be plotted for users to monitor coverage progress and data generation efficiency. However, the computational complexity increases rapidly with the magnitude of k, and the size and dimension of the dataset. A naïve algorithm that enumerates through all $n!/[k!\,(n!-k!)]$ data point combinations and finds all mutually covered regions is computationally inefficient. The overlap among the covered regions requires additional and complex computation.

Asudeh et al. [28] proposed using Voronoi diagram to reduce the computational complexity when calculating data coverage [29, 30]. Given two samples, $t_i$ and $t_j$, from dataset D, any point on a line $h(i, j) = \{q | \Delta(q, t_i) = \Delta(q, t_j)\}$ is equidistant to the two points. The half-space $h^+(i, j) = \{q | \Delta(q, t_i) \leq \Delta(q, t_j)\}$ includes $t_i$, and any point in this half-space is closer to $t_i$. A polygon $V(i) = \bigcap_{\forall i \neq j} h^+(i, j)$ is a Voronoi cell of sample i in which any point is closer to $t_i$ than other samples in D. In this way, the aggregation of all Voronoi cells is the Voronoi diagram of the first order. Similarly, for $k^{th}$ order Voronoi diagram, the k-nearest neighbors of any point in a Voronoi cell V(S) belong to S, where $S \in D$ and $|S| = k$. For an arbitrary value of k used in data coverage, a $k^{th}$ order Voronoi diagram can be constructed. To find the covered area, an algorithm only needs to enumerate through the Voronoi cells, and only computes the concurrently covered area by the associated k samples in S. This method does not suffer from overlap as the feature space has been segregated into Voronoi cells.

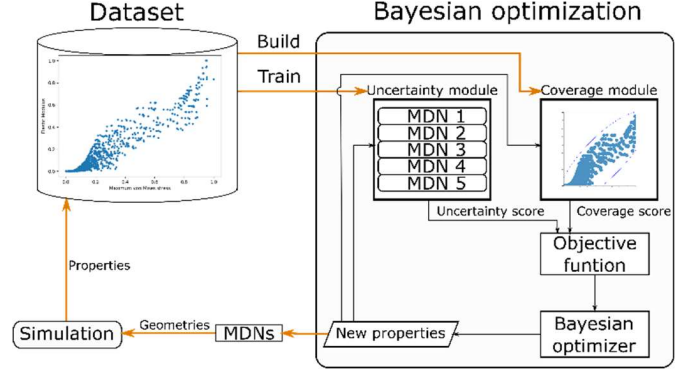

Figure 2: FairGen pipeline to iteratively generate missing properties in the property space.

Figure 3 demonstrates the use of Voronoi diagram to find the covered area by 1000 samples in the property space. Using the method proposed by Boots et al. [30], a $k^{th}$ order Voronoi diagram can be constructed in a time complexity of $O(k^2 n \log n)$ in a 2-dimensional space. For each Voronoi cell, the region covered by the associated data point is solved. The aggregation of all the regions is equivalent to the covered region by the dataset. Therefore, the area of the covered region is computed as the $S_C$ that reflects how well the property space is covered. $S_C - S'_C$ can be the metric to evaluate the selection of $D^P$, where $S_C$ and $S'_C$ are the coverage score after and before $D^P$ is added to the property space, respectively.

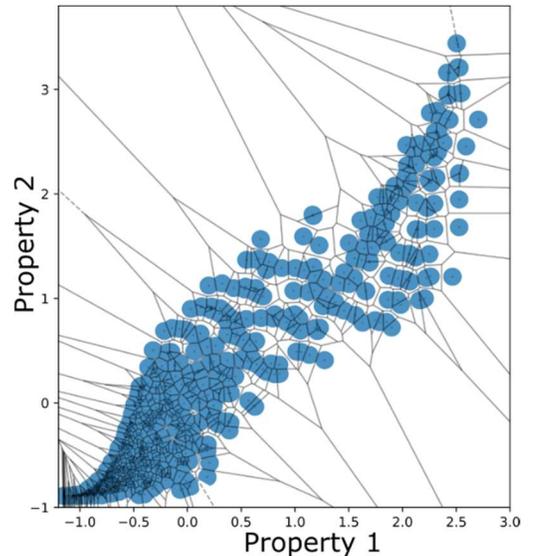

Figure 3: Data coverage using first order Voronoi diagram in a standardized 2-dimensional property space with k=1 and ρ=0.08.

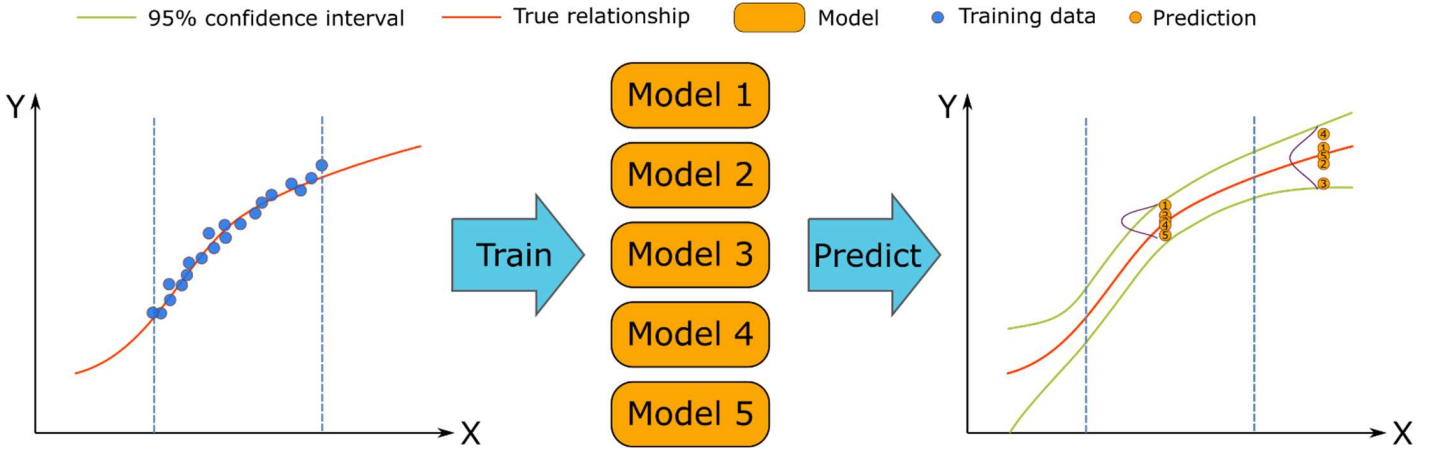

Figure 4: Modeling predictive uncertainty using ensemble method.

Through FairGen iterations and BO, the coverage module may consume considerable computational resources as Voronoi diagrams will be constructed repeatedly to compute new $S_C$'s. The advantage of Voronoi diagram is that a new diagram can be generated based on the preceding diagram to speed up the computation.

From an optimization perspective, the coverage improvement metric $S_C - S_C'$ can be simplified to $S_C$ during optimization since $S_C'$ is a constant. Moreover, $S_C$ as the objective function of BO encourages the selection of target properties that are far away from the existing properties. Taking those properties as the input, the MDN model will generate shapes that do not correspond to them. Thus, an uncertainty module is constructed to resolve this issue.

### 2.3. Uncertainty module

The uncertainty module calculates the predictive uncertainties of the MDN models for a given $D^P$. The predictive uncertainties form an uncertainty score ($S_U$) that penalizes the objective function to prevent selecting a $D^P$ that yields uncertain and thus potentially inaccurate shape predictions. There are two types of uncertainties: aleatoric and epistemic uncertainties [31]. Aleatoric uncertainty describes the inherent randomness such as sensor noise and measurement errors; epistemic uncertainty characterizes missing knowledge such as missing data or variables [32]. In such a context, the predictive uncertainty is to be modeled and utilized to guide BO.

Deep ensemble is a scalable and robust method to model predictive uncertainty [33]. To estimate the predictive uncertainty, multiple probabilistic neural network (NN) models are trained with different weight initialization and training data shuffling. The models are treated as a uniformly weighted mixture model where the predictions are combined as:

$$p(y|x) = M^{-1} \sum_{m=1}^{M} p_{\theta_m}(y|x, \theta_m) \quad (3)$$

where x is the input, y is the prediction, M is the number of models, and θ are the parameters. For regression problem, the prediction is a of Gaussian mixture:

$$M^{-1} \sum N(\mu_{\theta_m}(x), \sigma^2_{\theta_m}(x)) \quad (4)$$

where μ and $\sigma^2$ are the mean and variance, respectively. This mixture can be approximated as one Gaussian distribution where the mean and variance are:

$$\mu_*(x) = M^{-1} \sum_{m=1}^{M} \mu_{\theta_m}(x) \quad (5)$$

$$\sigma^2_*(x) = M^{-1} \sum_{m=1}^{M} \left( \sigma^2_{\theta_m}(x) - \mu^2_{\theta_m}(x) \right) - \mu_*(x) \quad (6)$$

Suppose the true relationship in Figure 4 is to be modeled with some training data collected. Given the same input, each model will provide a prediction, $y_m$, as a Gaussian distribution. The five predictions are approximated using one Gaussian distribution. If the input is within the region where data is available, the variance of the prediction is small, indicating small predictive uncertainty. If the input has no training data nearby, the variance of the predictions is large, characterizing a large predictive uncertainty. The deep ensemble method essentially investigates the difference among the M distributions learned by the M models. The same rationale is utilized to build the uncertainty module and obtain a $D^P$ with low predictive uncertainty through BO.

In generative design, generative models such as MDN are trained to predict the shapes that possess the input properties. MDNs proposed by Bishop [34] use the output discrete values from NNs to create a mixed Gaussian distribution and then, train the NNs to achieve consistency between the training dataset and the mixed distribution. Figure 5 depicts the structure of MDN comprised of a deep NN and a mixed Gaussian. The input layer receives the target properties. The output of the deep NN is reparametrized to construct a batch of Gaussian distributions, which are combined to form a Gaussian mixture. Design shapes are then sampled from the mixed Gaussian distribution. MDN is chosen to build the uncertainty module because it has embedded uncertainty measurement functionality. Thus, the deep ensemble method to characterize predictive uncertainty can be extended to MDN.

The previous deep ensemble scenario in Figure 4 describes a mixture of several single univariate Gaussian distributions. Modeling the predictive uncertainty of MDNs requires a mixture of several batches of multivariate Gaussian distributions (Figure 6 (a)). Each batch of Gaussian distributions is from one MDN model and each Gaussian distribution has d dimensions. Each model learns G distributions instead of one distribution in the previous example. Therefore, the deep ensemble method must be modified to investigate the difference among the M batches of G distributions learned by the M models. The assumption is that the M models are trained to learn the same G distributions, which characterize the true marginal distributions of the output variables.

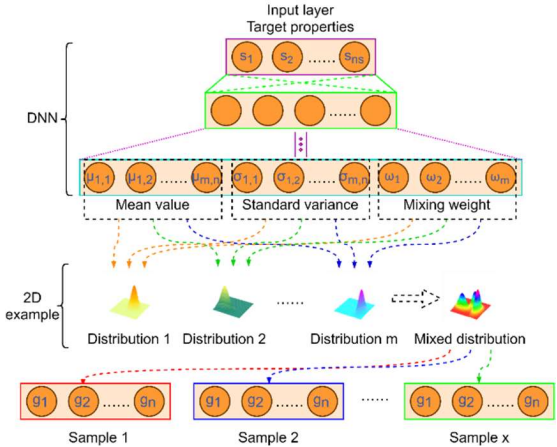

**Figure 5: Structure of an MDN.**

The first step is to find the correspondence of the G distributions from different MDNs using the training data. Although the MDNs learn the same ground truth distributions, the orders can be different. The approximation method using equations (5) and (6) must be conducted among the corresponding distributions indicated by the arrows in Figure 6 (a). When the training input X of size $n \times p$ is fed into an MDN, three matrices describing the proportions $(n \times G)$, means $(n \times G \times d)$, and variances $(n \times G \times d)$ of the G distributions will be the output. The corresponding distributions should have mean matrices close to each other. With this trait, the correspondence of distributions from multiple MDNs can be discovered by calculating the differences among the mean matrices.

After the correspondence is established, the corresponding distributions are approximated using one Gaussian distribution:

$$\mu_{*,g}(x) = M^{-1} \sum_{m=1}^{M} \mu_{\theta_{m,g}}(x) \qquad (7)$$

$$\sigma^2_{*,g}(x) = M^{-1} \sum_{m=1}^{M} \left( \sigma^2_{\theta_{m,g}}(x) - \mu^2_{\theta_{m,g}}(x) \right) - \mu_{*,g}(x) \qquad (8)$$

$$for \; \forall \; g = 1, 2, \dots, G$$

where each $\sigma^2_{*,g}(x)$ has a size of $1 \times d$. To obtain an uncertainty score that characterize the predictive uncertainty of the MDN models regarding a property input x, the variances are summed across G aggregated distributions and d dimensions:

$$S_U(x) = \sum_{g=1}^{G} \sigma^2_{*,g}(x) \times J_d \qquad (9)$$

where J is a $d \times 1$ matrix of ones.

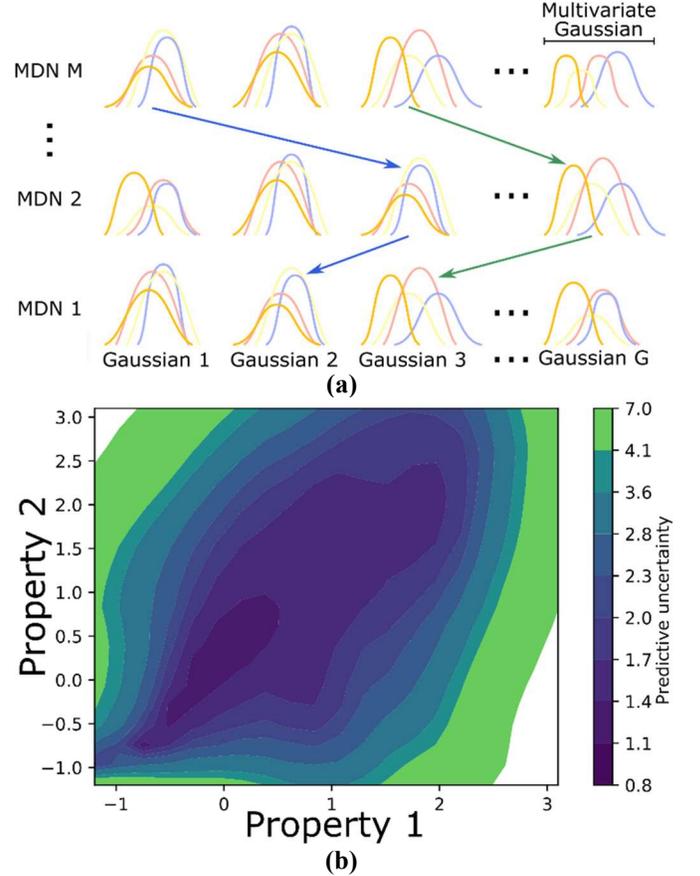

**Figure 6: Uncertainty module using deep ensemble method to model predictive uncertainty: a) Gaussian mixtures; and b) predictive uncertainty heatmap.**

Using the uncertainty module, the predictive uncertainty can be obtained at an arbitrary x in the property space. The example of a predictive uncertainty heatmap is plotted in Figure 6 (b). This heatmap indicates that the predictive uncertainty is low at the region where data is abundant and thus conveying sufficient knowledge to the model. As x travels toward the regions where data are sparse or absent, the predictive uncertainty increases, signaling a high potential for inaccurate predictions. Flexibility is the reason why the uncertainty score is used as a penalty instead of a constraint. By tuning the penalty factor (ψ), FairGen can switch between exploration and exploitation modes to actively search outside or within the sampled regions. Moreover, there will be fluctuation of the overall uncertainty levels, which could be compensated by the penalty factor.

This module might impose a high computational cost on the pipeline as multiple MDNs must be trained at every FairGen iteration. To speed up the uncertainty module, parallel training of the multiple models can be implemented as they are independent of each other. Transfer learning can help reduce training time. Instead of training from a randomly initialized model at every FairGen iteration, the models trained during the last iteration can be re-trained with the new dataset.

### 2.4. Bayesian optimization

The optimization function in FairGen finds the optimal $D^P = \{x_1, x_2, ..., x_{n_p}\}$ as the input to the generative models for design generation. At the $i^{th}$ iteration, the coverage and uncertainty modules calculate the coverage and uncertainty scores, accounting for the entire $D^P$:

$$S_C(D) = S_C(D^i \cup D^P) \quad (10)$$

$$S_U(D^P) = \sum_{i=1}^{n_p} \sum_{g=1}^{G} \sigma_{*,g}^2(x_i) \times J_d \quad (11)$$

The objective function of the BO can be formulated as:

$$\max_{D^P} f(D^P) = S_C(D) - \psi S_U(D^P) \quad (12)$$

After $D^P$ is determined by BO, the MDNs trained during the construction of the uncertainty module are utilized to generate design shapes. As $D^P$ is found with the penalty of their predictive uncertainties, some accurate estimations of the design shapes are likely to be obtained from the MDNs. Thereafter, the designs generated are analyzed using simulation to acquire the real properties. Finally, the shapes and properties generated during this iteration are added to the dataset. The next iteration can be executed with the updated dataset to further explore the property space.

## 3. RESULTS

This section exhibits the S-slot design case study with respect to the design problem, FairGen setting, and results.

### 3.1. S-slot design space exploration

In this paper, a case study of S-shaped perforated auxetic metamaterial design is conducted. S-slot designs have been proven to have an enhanced fatigue life due to its lower von Mise stress compared to the traditional circular design [35]. A dataset will be generated using FairGen and compared with conventional methods. The design spaces in this case study, including the shape and property spaces, are defined in this subsection. As shown in Figure 7, the S-slot is defined by four parameters including slot tail height (h), slot cap length (a), slot cap height (b) and cap rotation (α). The slot thickness, vertical spacing (VS) and horizontal spacing (HS) are fixed in this case study.

Maximum von Mises stress (MS) and EM are investigated in this DSE problem. As stress concentrations are the main reason for crack initiation, the MS of S-slot designs needs to be considered during the design process. Ideally, the MS in the design should be as small as possible. EM is also a mechanical property frequently discussed in research articles related to auxetic metamaterial [36]. The definition of optimal EM is determined based on the application as mentioned in the introduction. The goal of this case study is to generate a design dataset to build a generative model that predicts the design shapes given the required MS and EM. This dataset should efficiently explore the property space to possess abundant generative design knowledge. Although the design should have a small MS, the data generation process is not driven toward small MS regions to demonstrate a general case.

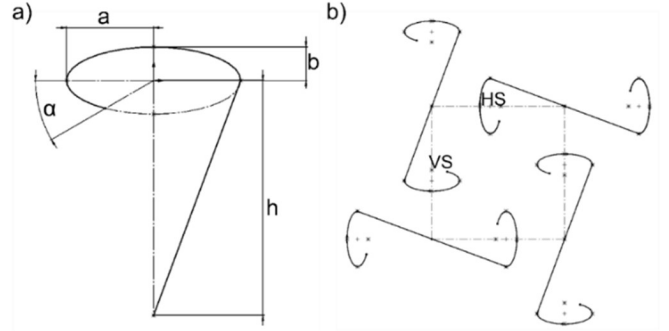

**Figure 7: S-slot design. a) Geometric parameters defining the S-slot; and b) Slot layout.**

We adopted the same numerical simulation as in the previous research [37] using static linear analysis with 3-dimensional triangle shell elements (S3R) on a unit cell with periodic boundary conditions in Abaqus to generate our simulation dataset. Although the elastic-plastic behavior is not considered, this simulation takes a relatively low computational cost and still provides stress distribution information related to crack initiation.

### 3.2. FairGen setting and iterations

The initial dataset consists of the shapes and properties of 1000 designs sampled using grid search from the shape space. The properties are standardized to the range of around [-1, 3] to facilitate the subsequent ML and FairGen operations. For the coverage module, ρ is 0.08 because a 2% percentage error of the property is acceptable in property prediction tasks. k is 1 because the initial dataset has only partially explored the property space. The uncertainty module includes 5 MDN models, which have six hidden layers, 10 Gaussian distributions, and 3000 training epochs. The uncertainty penalty is 0.1. BO will find the optimal 3 target properties in 50 iterations and 10 extra random walks. In this setting, it was found that selecting more than 3 target properties is likely to yield some unreasonable property selections. Experiments were run on a computer with a $12^{th}$ Gen Intel i7 processor with 16 gigabytes of available RAM on Windows 11. The models were trained in the central processing unit.

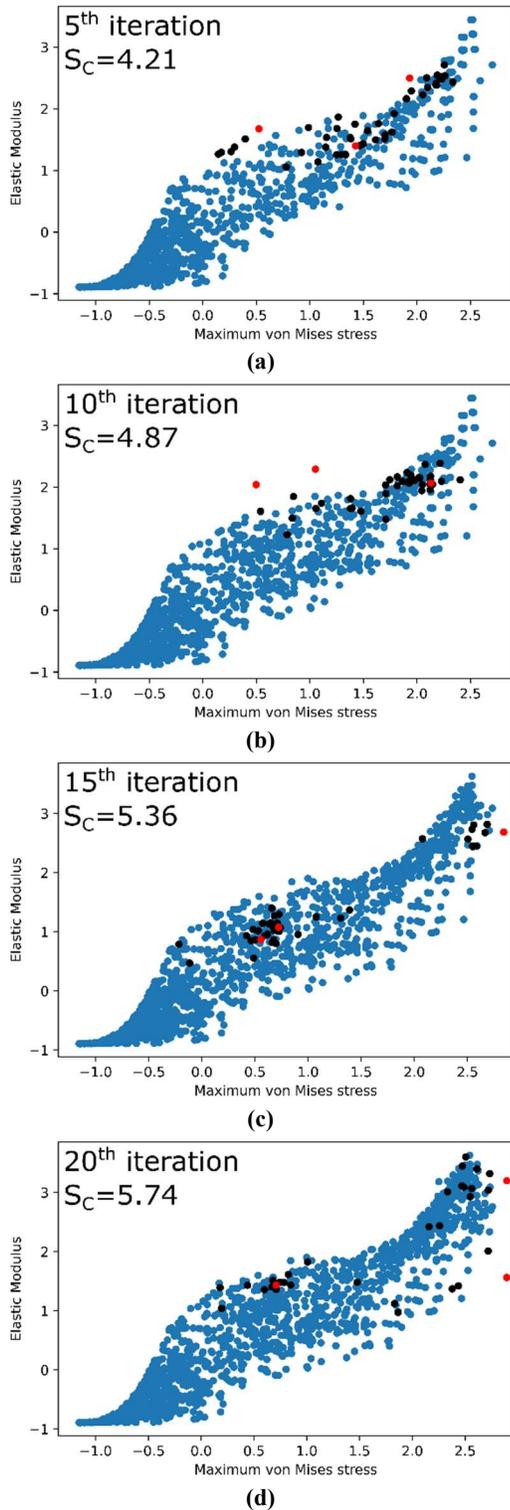

**Figure 8:** Iterative results of FairGen in the case study at a) 5$^{th}$ iteration with 1178 samples; b) 10$^{th}$ iteration with 1389 samples; c) 15$^{th}$ iteration with 1579 samples; and d) 20$^{th}$ iteration with 1769 samples

At the beginning of every FairGen iteration, the existing dataset was used to initialize the Voronoi diagram in the coverage module and train 5 MDNs in the uncertainty module. The two modules output the coverage and uncertainty scores for the $D^P$ selected at every BO iteration. The scores were combined to compute the objective function, which guides the Bayesian optimizer to select the $D^P$ for the next BO iteration. The final $D^P$ selected by BO both optimally increased the data coverage and yielded reasonable shape predictions. New designs were generated using the 5 MDNs trained in the uncertainty module with $D^P$ as the input. For each property in $D^P$, 3 designs were generated from each MDN, resulting in 45 new designs per FairGen iteration. Thereafter, the new designs were subject to manufacturability check to filter our infeasible designs such as S-slot intercept and thin wall. The properties of the feasible designs were obtained from simulation, and then added to the existing dataset. Some designs with properties that extended the coverage to the lower-right part of the properties space were regarded as outliers because they possess high maximum stress on the design. Such properties are undesirable and might bring some errors from the simulation.

Figure 8 showcases the properties of the generated designs at some FairGen iterations. At the 5$^{th}$ iteration, $S_U$ was increased from 3.5 at the beginning to 4.2 (Figure 8 (a)). One target property aimed to exploit a void within the sampled region. The generated properties successfully filled the void. Two target properties tried to explore the uncovered region. Many new designs were generated that considerably expanded the sampled region. At the 10$^{th}$ iteration, one target property exploited a void and densified the surrounding region (Figure 8 (b)). The other two target properties led to the finding of two designs that expanded the sampled region. At the 15$^{th}$ iteration, two target properties exploited the sampled region and one target property explored the rightmost unexplored region (Figure 8 (c)). At the 20$^{th}$ iteration, one target property searched a void region, and two properties explored the rightmost region (Figure 8 (d)).

## 4. DISCUSSION

After 20 FairGen iterations, 799 new designs have been generated based on the 1000 initial designs. To form a comparison and investigate the effectiveness of FairGen, 3000 designs were generated using grid sampling and randomized sampling from the shape space, respectively. The former is a conventional DOE method with a strong bias toward the designated geometrical parameters [38]. The latter utilizes a Latin Hypercube sampling (LHS) that encourages shape diversity [39]. The comparison among the three sampling methods will be analyzed with respect to the data coverage, property space exploration, generative modeling, and computational cost.

### 4.1. Data coverage and property space exploration

Figure 9 (a) reveals the increase in data coverage as the number of samples increased using the three sampling techniques. Grid sampling started from a low coverage score than randomized sampling because of its strong bias. FairGen started from the same coverage score as grid sampling because it was initialized with a dataset based on grid sampling. Although randomized sampling offered a high initial coverage score, data coverage is increasing at the same speed as grid sampling. Also, they both show the trend to converge. On the contrary, the FairGen coverage score curve has not shown the trend to converge. Using FairGen, the data coverage was rapidly improved and quickly surpassed randomized sampling at the second iteration. Eventually, FairGen sampling reached a coverage score of 5.8 while the other two methods were below 4.8.

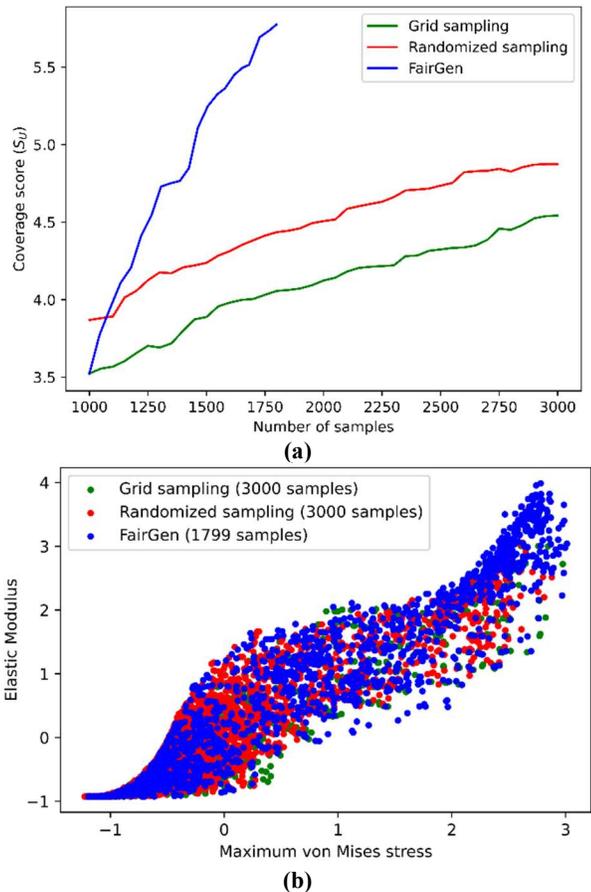

Figure 9: Comparison among FairGen, grid sampling, and randomized sampling with respect to a) data coverage (k=1 and ρ=0.08); and b) property space exploration.

Figure 9 (b) visualizes the datasets generated by the three methods. Grid sampling provided the worst property space exploration effect as most of its samples are covered by the other two methods. Grid sampling intensively sampled the low-MS and low-EM region, while the rest of the property space is either sparsely populated or unexplored. Samples were likely to stick together and repetitively cover a region. Randomized sampling also intensively searched the low-MS and low-EM region, which was less severe than grid sampling. Samples were likely to evenly disperse instead of forming blocks, but also created some greater voids. FairGen significantly avoided the intensive search effect and generated more evenly distributed properties. It almost established the full contour of the sampled area with only a small portion established by others. In reality, the best design at a certain level of elastic modulus should possess the smallest maximum von Mises stress. Figure 9 (b) indicates that FairGen offered the smallest maximum von Mises stress at almost all elastic modulus levels with fewer samples, especially at high elastic moduli. In conclusion, FairGen has a better capability to explore and exploit the potential of the design in DSE.

### 4.2. Generative modeling

The purpose of increasing data coverage is to improve the performance of the ML models. This subsection investigates the effect of FairGen on generative models. MDN models were trained using the dataset acquired from the three sampling techniques. 50 test properties were randomly sampled within the sampled region of the property space. For each test property, each MDN predicted 10 shapes. In total, each MDN predicted 500 shapes, whose properties were analyzed using simulation. The real properties were compared with the target properties to find the predictive errors. To avoid being misled by randomness, tests were conducted at different data sizes: 1200, 1400, 1600, and 1800 designs in the training set (Table 1). This way, both the predictive error and the trend can be the evidence for comparison.

The mean absolute error (MAE) of generative model predictions can sometimes be misleading for generative models as some outliers might be generated. Thus, both the MAEs (Table 1) and the absolute prediction error scatter plots (Figure 10) are provided. The horizonal and vertical axes in Figure 10 represent the absolute prediction errors of MS and EM, respectively. Table 1 indicates that all three sampling techniques helped reduce the MAE as the number of samples increased. The MAEs of FairGen were always 1/3 smaller than grid sampling and on average 1/8 smaller than randomized sampling. This could be verified by the scatter plots. When trained with 1200 designs (Figure 10 (a)), large prediction errors were obtained from all models. The performances of FairGen and randomized sampling are close to each other and are significantly better than grid sampling. As more designs being generated, the prediction errors of the three methods became smaller and smaller. Meanwhile, the models trained using FairGen generated datasets performed better than the models trained using randomly sampled datasets (Figure 10 (b)-(d)). The generative modeling test results revealed that data generated using FairGen efficiently explored the property space to embed more knowledge regarding generative design.

Table 1: MAEs of the MDNs trained with different numbers of training examples generated from FairGen, grid sampling, and randomized sampling.

| n | FairGen | Grid sampling | Randomized sampling |
|---|---|---|---|
| 1200 | 0.2067 | 0.2989 | 0.1835 |
| 1400 | 0.1499 | 0.2095 | 0.1610 |
| 1600 | 0.1408 | 0.2081 | 0.1671 |
| 1800 | 0.1286 | 0.1903 | 0.1574 |

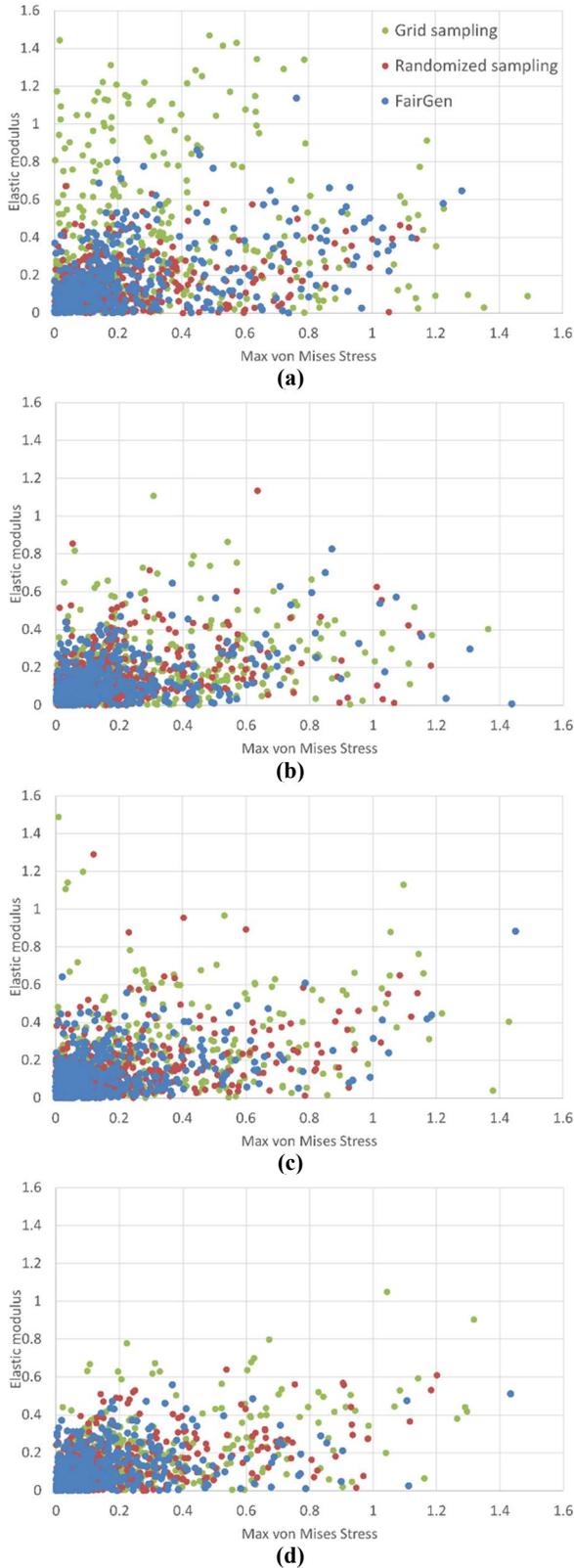

**Figure 10: Scatter plots of the absolute prediction errors yielded by different sampling techniques at a) n=1200; b) n=1400; c) n=1600; and d) n=1800.**

### 4.3. Computational cost

The goal of FairGen is to reduce the time and resources required to build an unbiased dataset. It has been shown that FairGen provides higher data coverage and better generative modeling capabilities. Nonetheless, it also adds the coverage module, uncertainty module, and BO to the data generation pipeline. The upper bound of the time complexity of Voronoi diagram construction is $O((d+1)n^{d/2} k^{d/2+1})$ [40]. For a 2-dimensional Voronoi diagram where $k=1$, the cost can be reduced to $O(n \log n)$ [22]. The number of Voronoi cells is bounded by $O(n^{d/2} k^{d/2})$, which yields an upperbound of n cells in this case study [28]. For each cell, the complexity of identifying the covered region is $O(k(d+1))$. Thus, the complexity to compute the entire covered area is bounded by $O((d+1)n^{d/2} k^{d/2+1})$, which is $O(3n)$ in this case study. The computational cost of building the uncertainty module is equivalent to training 5 MDNs. At every FairGen iteration, the two modules must be constructed again. At every BO iteration, the Voronoi diagram covered area, and the output of 5 MDNs are computed. This case study has relatively small n, k, and d such that the baseline pipeline is not computationally heavy. For large values of n, k and d, methods such as transfer learning [41] and data coverage approximation [28] can be utilized to significantly reduce the computational cost.

For the computation unit in this case study, it took around 37 seconds to complete the simulation of one design. From 1000 to 1799 samples, the time to initialize the coverage module ranged from 2 to 3 seconds. The time to build the uncertainty module increased from 50 to 78 seconds. The entire BO consumed around 120 to 240 seconds. The computational time spent on 20 FairGen iterations was around 4960 seconds. The time to generate 799 designs using FairGen is equivalent to 933 designs generated by geometric sampling techniques. With reasonable extra computational time, FairGen achieved exceptional property exploration and generative modeling results.

### 5. CONCLUSIONS

This paper proposed and demonstrated the FairGen pipeline that efficiently explores the property space in DSE problems. The existing methods cannot directly generate missing properties in a design dataset to explore the potential of the design. This leads to missing knowledge and unsatisfactory ML model performance. FairGen finds the missing properties and actively generates designs that provide those properties to complement the dataset. Its coverage module detects unexplored regions in the property space using a fairness metric. The uncertainty module evaluates the predictive uncertainty in the property space to avoid sampling from the regions about which the generative models are uncertain. BO integrates the coverage and uncertainty modules to solve for the target properties that both maximally increase the data coverage and yield reasonable shape predictions. Thereafter, the target properties are input into the generative models to generate the associated shapes, whose properties are analyzed using simulation. The new designs are added to the dataset and the above steps can be implemented iteratively to improve data coverage.

In the S-slot case study, FairGen was implemented to investigate its efficiency, starting with a dataset that has 1000 designs sampled using grid geometric sampling. After 20 iterations, 799 new designs were generated. The coverage score was increased from 3.5 to 5.8 whereas grid sampling and randomized sampling could only increase the coverage score to 4.8 at 3000 samples. FairGen also significantly expanded the sampled region in the property space more than the other sampling techniques. The expanded area means designs with better properties can be obtained from the dataset generated using FairGen. The generative modeling test revealed that the models trained using FairGen generated dataset reduced the MAE by 1/3 and 1/8 on average compared with the datasets generated using grid sampling and randomized sampling, respectively. Computationally, the time spent on generating 799 designs using baseline FairGen is equivalent to generating 933 designs using other sampling methods in the current setting of the case study and computational resources.

The limitation of FairGen is the lack of a shape diversity mechanism. Future work will focus on the simultaneous improvement of shape and property fairness. Moreover, FairGen can be modified to actively drive data generation toward desirable property regions.

## ACKNOWLEDGEMENTS

This work is funded by McGill University Graduate Excellence Fellowship Award [grant number 00157]; Mitacs Accelerate program [grant number IT13369]; and McGill Engineering Doctoral Award (MEDA).

## DECLARATION OF COMPETING INTEREST

The authors declare that they have no known competing interests.

## REFERENCES

[1] Yan, Wentao, Lin, Stephen, Kafka, Orion L, Lian, Yanping, Yu, Cheng, Liu, Zeliang, Yan, Jinhui, Wolff, Sarah, Wu, Hao, and Ndip-Agbor, Ebot. "Data-driven multi-scale multi-physics models to derive process–structure–property relationships for additive manufacturing." *Computational Mechanics* Vol. 61 (2018): pp. 521-541.

[2] Pilarski, Sebastian, Staniszewski, Martin, Villeneuve, Frederic, and Varro, Daniel. "On artificial intelligence for simulation and design space exploration in gas turbine design." *2019 ACM/IEEE 22nd International Conference on Model Driven Engineering Languages and Systems Companion (MODELS-C)*. pp. 170-174. 2019.

[3] Chen, Wei and Ahmed, Faez. "Padgan: Learning to generate high-quality novel designs." *Journal of Mechanical Design* Vol. 143 No. 3 (2021).

[4] Jang, Seowoo, Yoo, Soyoung, and Kang, Namwoo. "Generative design by reinforcement learning: enhancing the diversity of topology optimization designs." *Computer-Aided Design* Vol. 146 (2022): pp. 103225.

[5] Agrawal, Akash and McComb, Christopher. "Reinforcement Learning for Efficient Design Space Exploration With Variable Fidelity Analysis Models." *Journal of Computing and Information Science in Engineering* Vol. 23 No. 4 (2023): pp. 041004.

[6] Wang, Liwei, Chan, Yu-Chin, Liu, Zhao, Zhu, Ping, and Chen, Wei. "Data-driven metamaterial design with Laplace-Beltrami spectrum as "shape-DNA"." *Structural and multidisciplinary optimization* Vol. 61 (2020): pp. 2613-2628.

[7] Sun, Hongbo and Ma, Ling. "Generative design by using exploration approaches of reinforcement learning in density-based structural topology optimization." *Designs* Vol. 4 No. 2 (2020): pp. 10.

[8] Oh, Sangeun, Jung, Yongsu, Kim, Seongsin, Lee, Ikjin, and Kang, Namwoo. "Deep generative design: Integration of topology optimization and generative models." *Journal of Mechanical Design* Vol. 141 No. 11 (2019).

[9] Ling, Chunyan, Kuo, Way, and Xie, Min. "An overview of adaptive-surrogate-model-assisted methods for reliability-based design optimization." *IEEE Transactions on Reliability* (2022).

[10] Nakamura, Gen and Potthast, Roland. *Inverse modeling*. IOP Publishing, (2015).

[11] Suresh, Harini and Guttag, John. "A framework for understanding sources of harm throughout the machine learning life cycle." *Equity and access in algorithms, mechanisms, and optimization*. (2021): p. 1-9.

[12] Lee, Doksoo, Chan, Yu-Chin, Chen, Wei, Wang, Liwei, and Chen, Wei. "t-METASET: Task-Aware Generation of Metamaterial Datasets by Diversity-Based Active Learning." *International Design Engineering Technical Conferences and Computers and Information in Engineering Conference*. pp. V03AT03A011. 2022.

[13] Chan, Yu-Chin, Ahmed, Faez, Wang, Liwei, and Chen, Wei. "METASET: Exploring shape and property spaces for data-driven metamaterials design." *Journal of Mechanical Design* Vol. 143 No. 3 (2021).

[14] Shahbazi, Nima, Lin, Yin, Asudeh, Abolfazl, and Jagadish, HV. "A Survey on Techniques for Identifying and Resolving Representation Bias in Data." *arXiv preprint arXiv:2203.11852* (2022).

[15] Haixiang, Guo, Yijing, Li, Shang, Jennifer, Mingyun, Gu, Yuanyue, Huang, and Bing, Gong. "Learning from class-imbalanced data: Review of methods and applications." *Expert systems with applications* Vol. 73 (2017): pp. 220-239.

[16] Branco, Paula, Torgo, Luís, and Ribeiro, Rita P. "A survey of predictive modeling on imbalanced domains." *ACM computing surveys (CSUR)* Vol. 49 No. 2 (2016): pp. 1-50.

[17] Chen, Wei and Ahmed, Faez. "Mo-padgan: Reparameterizing engineering designs for augmented multi-objective optimization." *Applied Soft Computing* Vol. 113 (2021): pp. 107909.

[18] Nobari, Amin Heyrani, Chen, Wei, and Ahmed, Faez. "Pcdgan: A continuous conditional diverse generative adversarial network for inverse design." *arXiv preprint arXiv:2106.03620* (2021).

[19] Tan, Yew Teck, Kunapareddy, Abhinav, and Kobilarov, Marin. "Gaussian process adaptive sampling using the cross-entropy method for environmental sensing and monitoring." *2018 IEEE International Conference on Robotics and Automation (ICRA)*. pp. 6220-6227. 2018.

[20] Zhang, Qi, Wu, Yizhong, Lu, Li, and Qiao, Ping. "An adaptive dendrite-HDMR metamodeling technique for high-dimensional problems." *Journal of Mechanical Design* Vol. 144 No. 8 (2022): pp. 081701.

[21] Kapusuzoglu, Berkcan, Mahadevan, Sankaran, Matsumoto, Shunsaku, Miyagi, Yoshitomo, and Watanabe, Daigo. "Adaptive surrogate modeling for high-dimensional spatio-temporal output." *Structural and Multidisciplinary Optimization* Vol. 65 No. 10 (2022): pp. 300.

[22] Sun, Qi, Bai, Chen, Geng, Hao, and Yu, Bei. "Deep neural network hardware deployment optimization via advanced active learning." *2021 Design, Automation & Test in Europe Conference & Exhibition (DATE)*. pp. 1510-1515. 2021.

[23] Wang, Yi, Franzon, Paul D, Smart, David, and Swahn, Brian. "Multi-fidelity surrogate-based optimization for electromagnetic simulation acceleration." *ACM Transactions on Design Automation of Electronic Systems (TODAES)* Vol. 25 No. 5 (2020): pp. 1-21.

[24] Kolesnikov, VI, Pashkov, DM, Belyak, OA, Guda, AA, Danilchenko, SA, Manturov, DS, Novikov, ES, Kudryakov, OV, Guda, SA, and Soldatov, AV. "Design of double layer protective coatings: Finite element modeling and machine learning approximations." *Acta Astronautica* Vol. 204 (2023): pp. 869-877.

[25] Xu, Yanwen, Zheng, Zhuoyuan, Arora, Kanika, Senesky, Debbie G, and Wang, Pingfeng. "Hall Effect Sensor Design Optimization With Multi-Physics Informed Gaussian Process Modeling." *International Design Engineering Technical Conferences and Computers and Information in Engineering Conference*. pp. V03BT03A028. 2022.

[26] Liu, Zheng, Renteria, Anabel, Zheng, Zhuoyuan, Wang, Pingfeng, and Li, Yumeng. "Design of Additively Manufactured Functionally Graded Cellular Structures." *IIE Annual Conference. Proceedings*. pp. 1-6. 2022.

[27] Catania, Barbara, Guerrini, Giovanna, and Accinelli, Chiara. "Fairness & friends in the data science era." *AI & SOCIETY* (2022): pp. 1-11.

[28] Asudeh, Abolfazl, Shahbazi, Nima, Jin, Zhongjun, and Jagadish, HV. "Identifying insufficient data coverage for ordinal continuous-valued attributes." *Proceedings of the 2021 international conference on management of data*. pp. 129-141. 2021.

[29] Aurenhammer, Franz. "Voronoi diagrams—a survey of a fundamental geometric data structure." *ACM Computing Surveys (CSUR)* Vol. 23 No. 3 (1991): pp. 345-405.

[30] Boots, Barry, Sugihara, Kokichi, Chiu, Sung Nok, and Okabe, Atsuyuki. "Spatial tessellations: concepts and applications of Voronoi diagrams." (2009).

[31] Hora, Stephen C. "Aleatory and epistemic uncertainty in probability elicitation with an example from hazardous waste management." *Reliability Engineering & System Safety* Vol. 54 No. 2-3 (1996): pp. 217-223.

[32] Hüllermeier, Eyke and Waegeman, Willem. "Aleatoric and epistemic uncertainty in machine learning: An introduction to concepts and methods." *Machine Learning* Vol. 110 (2021): pp. 457-506.

[33] Lakshminarayanan, Balaji, Pritzel, Alexander, and Blundell, Charles. "Simple and scalable predictive uncertainty estimation using deep ensembles." *Advances in neural information processing systems* Vol. 30 (2017).

[34] Bishop, Christopher M, *Mixture density networks*. 1994, Aston University.

[35] Javid, Farhad, Liu, Jia, Rafsanjani, Ahmad, Schaenzer, Megan, Pham, Minh Quan, Backman, David, Yandt, Scott, Innes, Matthew C., Booth-Morrison, Christopher, Gerendas, Miklos, Scarinci, Thomas, Shanian, Ali, and Bertoldi, Katia. "On the design of porous structures with enhanced fatigue life." *Extreme Mechanics Letters* Vol. 16 (2017): pp. 13-17. https://doi.org/10.1016/j.eml.2017.08.002.

[36] Saxena, Krishna Kumar, Das, Raj, and Calius, Emilio P. "Three Decades of Auxetics Research − Materials with Negative Poisson's Ratio: A Review." *Advanced Engineering Materials* Vol. 18 No. 11 (2016): pp. 1847-1870. https://doi.org/10.1002/adem.201600053.

[37] Javid, Farhad, Liu, Jia, Rafsanjani, Ahmad, Schaenzer, Megan, Pham, Minh Quan, Backman, David, Yandt, Scott, Innes, Matthew C, Booth-Morrison, Christopher, and Gerendas, Miklos. "On the design of porous structures with enhanced fatigue life." *Extreme Mechanics Letters* Vol. 16 (2017): pp. 13-17.

[38] Lawrence, Patrick G, Roper, Wayne, Morris, Thomas F, and Guillard, Karl. "Guiding soil sampling strategies using classical and spatial statistics: A review." *Agronomy Journal* Vol. 112 No. 1 (2020): pp. 493-510.

[39] Zhang, Feng, Cheng, Lei, Wu, Mingying, Xu, Xiayu, Wang, Pengcheng, and Liu, Zhongbing. "Performance analysis of two-stage thermoelectric generator model based on Latin hypercube sampling." *Energy Conversion and Management* Vol. 221 (2020): pp. 113159.

[40] Edelsbrunner, Herbert and Seidel, Raimund. "Voronoi diagrams and arrangements." *Proceedings of the first annual symposium on Computational geometry*. pp. 251-262. 1985.

[41] Goodfellow, Ian, Bengio, Yoshua, and Courville, Aaron. *Deep learning*. MIT press, (2016).